\documentclass{article}

\usepackage{arxiv}

\usepackage[utf8]{inputenc} 
\usepackage[T1]{fontenc}    
\usepackage{hyperref}       
\usepackage{url}            
\usepackage{booktabs}       
\usepackage{amsfonts}       
\usepackage{nicefrac}       
\usepackage{microtype}      
\usepackage{lipsum}		
\usepackage{graphicx}
\usepackage{natbib}
\usepackage{doi}
\usepackage{algorithm}
\usepackage{algorithmic}

\usepackage{comment}   
\usepackage{multirow}
\usepackage{booktabs}
\usepackage{enumitem,amssymb}
\usepackage{subfigure}   
\usepackage{lipsum}
\usepackage{color,soul}
\makeatletter
\def\mathcolor#1#{\@mathcolor{#1}}
\def\@mathcolor#1#2#3{%
  \protect\leavevmode
  \begingroup
    \color#1{#2}#3%
  \endgroup
}
\makeatother

\usepackage{float}
\usepackage{amsmath,amsfonts,amsthm}
\usepackage{hyperref}
\usepackage{bm}
\title{Disentangling semantic features of macromolecules in Cryo-Electron Tomography}
   
\author{Kai Yi\\
 Department of Computer Science\\
 King Abdullah University of Science and Technology\\
kai.yi@kaust.edu.sa
 \And
 Jianye Pang\\
  Department of Computer Science\\ 
 Xi'an Jiaotong University\\
jianye.pang97@gmail.com
 \And 
 Yungeng Zhang\\
 Department of Computer Science\\
 Peking University\\
zhangyungeng@pku.edu.cn
 \And
 Xiangrui Zeng\\
  Computational Biology Department\\ 
 Carnegie Mellon University\\
xiangruz@andrew.cmu.edu\\
 \And
  Min Xu\\
  Computational Biology Department\\ 
 Carnegie Mellon University\\
 mxu1@cs.cmu.edu\\
}

\renewcommand{\shorttitle}{3D-SpVAE, Kai Yi. \textit{et al}}

\hypersetup{
pdftitle={3D-SpVAE},
pdfsubject={CS},
pdfauthor={Kai Yi},
}

\begin{document}
\maketitle

\begin{abstract}
Cryo-electron tomography (Cryo-ET) is a 3D imaging technique that enables the systemic study of shape, abundance, and distribution of macromolecular structures in single cells in near-atomic resolution. However, the systematic and efficient \emph{de novo} recognition and recovery of macromolecular structures captured by Cryo-ET are very challenging due to the structural complexity and imaging limits. Even macromolecules with identical structures have various appearances due to different orientations and imaging limits, such as noise and the missing wedge effect. Explicitly disentangling the semantic features of macromolecules is crucial for performing several downstream analyses on the macromolecules. This paper has addressed the problem by proposing a 3D Spatial Variational Autoencoder that explicitly disentangle the structure, orientation, and shift of macromolecules. Extensive experiments on both synthesized and real cryo-ET datasets and cross-domain evaluations demonstrate the efficacy of our method. 
\end{abstract}

\section{Introduction}
Cryo-electron tomography (cryo-ET) enables the 3D visualization of the structure and spatial organization of large macromolecules and their spatial interactions with other cellular components in single cells at sub-molecular resolution in near-native state \cite{li2021cryo}. It is the most important technique for studying macromolecular structures \emph{in situ}. However, the recognition and recovery of macromolecular structures captured by cryo-ET 3D images are extremely challenging due to structural complexity and imaging limits. From the structural point of view, the majority of the native macromolecular structures are unknown and very diverse \cite{hanson2010unknown, looso2010advanced}. The macromolecules may be inside a crowded environment\cite{ellis2001macromolecular}. From the imaging point of view, the cryo-ET tomograms are often of low signal-to-noise ratio (SNR) due to limited electron dose \cite{li2021cryo}. The tomograms also have missing values (a.k.a missing wedge effect) due to limited tilt angle range \cite{li2021cryo}. Therefore, to systematically recognize the macromolecular structures, the macromolecules must be separated into structurally homogeneous groups through subtomogram classification. A \emph{subtomogram} is a cubic subvolume of a tomogram that is likely to contain a single macromolecule. Due to such imaging limits as low SNR and missing wedge effect, a large number of subtomograms with identical structure and orientation must be averaged to recover macromolecular structure. However, the macromolecules of even identical structure have various orientations.

Because the majority of the native macromolecular structures are unknown\cite{hanson2010unknown, looso2010advanced}, the usage of template matching\cite{bohm2000toward} based recognition methods is limited. To recognize unknown macromolecular structures \emph{de novo},  template-free classification methods are needed. Existing template-free approaches focus on subtomogram alignments\cite{forster2008classification,bartesaghi2008classification,xu2012high-throughput,chen2013fast} and clustering-based classification\cite{bartesaghi2008classification}. However, the classification and alignment are often processed in separate steps. Traditional geometric based classification approaches\cite{bartesaghi2008classification} iteratively cluster and align subtomograms. They have very limited scalability and cannot process the millions of macromolecules captured by current electron microscopes. On the other hand, despite substantially higher scalability, deep learning based approaches\cite{xu2017deep} generally require supervision. However, the supervised deep learning-based approaches cannot be directly used to discover unknown structures, which significantly limited their usage. To solve this problem, an unsupervised autoencoder approach\cite{zeng2018convolutional} was developed. However, it relies on a separate pose normalization step with very limited ability to reduce orientation difference.

Considering the above limits of the cryo-ET data and current methods, we believe it is ideal to have an end-to-end unsupervised deep learning method that can simultaneously separate structures, orientations, and translations. In this work, we specifically focus on the following problem: \textit{How to learn the generative features reflecting specific characteristics of input 3D data?} To be more precise, given a subtomogram, can we build a model-based network that can summarize the input data as a set of latent generative vectors? In addition, can we learn transformation factors such as translations and rotations independently from the structures? We seek the answer to those questions from generative models.

We notice generative models gain a lot of insights into learning template-free patterns in an unsupervised manner. To learn transformation factors explicitly, we proposed several methods. It is the first time to learn disentangled semantic features from Cryo-ET to the best of our knowledge. We compare our proposed method with the generative implementation of Vanilla VAE in three forms. We investigate our method on both synthetic datasets and real-world datasets to verify the actual performance. For synthetic datasets, we want to determine whether it can represent some features of real data and find the most tolerant SNR of different models, which means the point with the best trade-off between SNR and performance (i.e., ELBO in our setting). We also demonstrate the performance and conduct an ablation study on real datasets. The results show our method is robust and can disentangle transformation factors on either synthetic datasets and real datasets.

Our paper is organized as follows. We summarize the works most related to us in different aspects in Section 2. Then we introduce our method and explain it thoroughly in Section 3. Next, we talk about the experimental setting and discussion the results in Section 4. Finally, we conclude the whole paper in Section 5.

\section{Related Work}

\subsection{Unsupervised Learning with VAEs}
Deep generative models are prevailing in solving a wide range of tasks. VAEs have been very successful for unsupervised representation learning of images, both realistic images and synthesis images~\cite{higgins2017beta, kim2018disentangling, chen2018isolating}. 

One of the earliest works is $\beta-VAE$\cite{higgins2017beta}, which uses a hyperparameter $\beta$ to balance the reconstruction and disentangled representative ability. Further, $Factor VAE$\cite{kim2018disentangling} modifies the routine of $\beta-VAE$ and proposes a new total correlation (TC) to improve the disentangling ability of VAE. TC is used to measure the dependence between variables. Inspired by them, $\beta-TCVAE$\cite{chen2018isolating} specifies why $\beta-VAE$ could work and what the KL-divergence represents, whose separated result contains index-code mutual information, TC, and dimension-wise KL. Another notable work related to ours is HRIC-VAE\cite{lopez2018information}, which employs the d-variable Hilbert-Schmidt Independence Criterion (dHSIC) to enforce independence between the latent factors. 

However, the methods mentioned above can not disentangle the hidden features explicitly. We can not say for sure if one hidden vector is encoded to represent a corresponding object characteristic. The method most related to us is \cite{bepler2019explicitly}, which isolates pose variables such as rotation and translation from 2D images. The difference of our approach compared to \cite{bepler2019explicitly} is that we try to encode the transformation factors independently from the 3D semantic content, which allows more adaptive pose orientations. 

\subsection{3D Disentangled Representation}       
Disentangled representation of 3D data is a critical topic. However, there are only a few methods proposed in this area~\cite{zhu2018visual, jiang2019disentangled}. \cite{zhu2018visual} proposed a GAN-based generative network to synthesize realistic images of objects with a 3D disentangled representation. This paper mainly focuses on shape, viewpoint, and texture. Different from that, our method pays attention to rotations and translations, which are more dynamic. Another notable method is in \cite{jiang2019disentangled}. They disentangle the hidden features of 3D face shape into two branches: the identity and expression branches. After extracting meaningful contexts, they fuse two parts. Moreover, the mentioned two approaches can not represent the disentangled factors from semantic content explicitly. Our proposed method cares explicitly about disentangling the transformation factors. 

\begin{figure*}
    \centering
    \includegraphics[width=1.0\textwidth, trim={70 80 80 80}, clip]{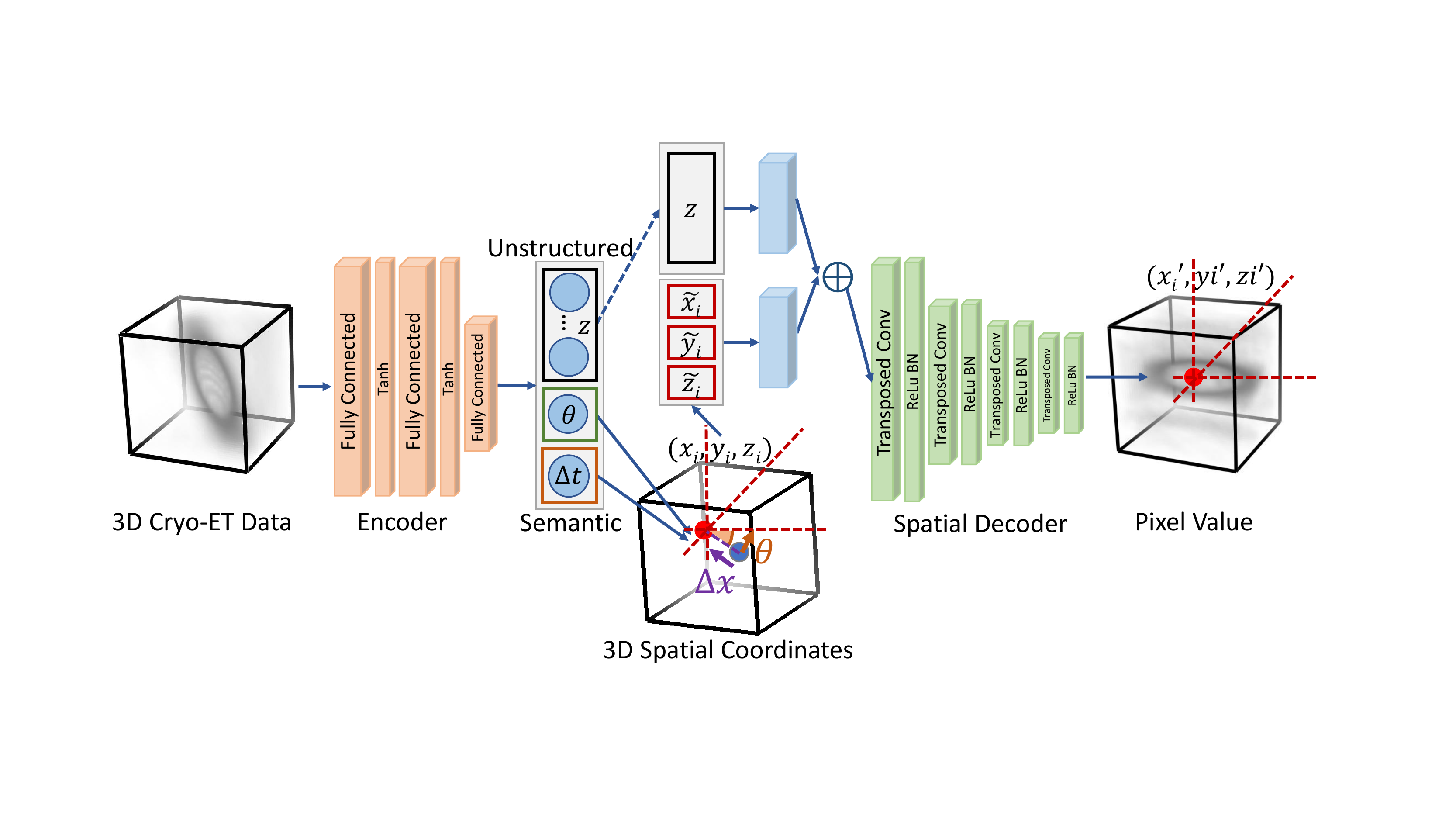}
    \vspace{-8mm}
    \caption{The network architecture of our proposed 3D-SpVAE. Our encoder first learns disentangled representations, which includes rotation factors $\theta$, translation factors $\Delta t$ and unstructured latent vectors $z$. Then we adopt a spatial coordinates transformer getting the converted coordinates of every pixel $\mathbf{x}_i = (x_{i}, y_{i}, z_{i})$ from rotation and translation factors. Finally we obtain the pixel value modeled by the probability at position $\mathbf{x}_i$ trained by adding feature vectors from two MLPs, the one is an unstructured mapping on $z$ and the other one is a coordinate-based semantic mapping on transformed coordinates $\mathbf{x}_{i}^{t} = (\tilde{x_{i}}, \tilde{y_{i}}, \tilde{z_{i}})$ to the spatial decoder. The encoder and the decoder can be either modeled as multiple layers perceptron or convolutional networks in our setting. Here we use block to represent fully connected layers (cuticolor), unstructured mapping and coordinate-based semantic mapping layers (blue), transposed convolution layers (green).}\label{fig:spatial-vae-arch}
\end{figure*}       

\section{Methods}
We aim to encode the transformation factor independently from the semantic content. To this end, we employ the 3D-SpVAE method to learn latent representations that disentangle volume rotation and translation from the content. We first elaborate on the general architecture of our proposed 3D-SpVAE. Then we present two essential components: the inference network, which learns the semantic representations and unstructured latent factors, and the 3D Spatial generator network, predicting the voxel-wise probability from the disentangled semantic features. Lastly, we proposed the corresponding variational lower-bound loss to minimize the whole model.
          
\subsection{Architecture of 3D-SpVAE}

Different from the standard VAE procedure, in the 3D spatial-VAE framework, the transformation parameters and other semantic factors are encoded, respectively, and a novel spatial generator module is employed to displace the standard VAE decoder. The network architecture can be found at Fig. \ref{fig:spatial-vae-arch}. Our method contains a standard encoder and a well-designed spatial decoder, which predict the pixel-level probability of a particular point instead of treating the pixels in an image as a whole. We denote the MLP implemented spatial decoder as 3D-SpVAE and the well-designed CNN implementation as 3D-SpVAE++. We attached the model architecture in the supplementary.

\subsection{3D Inference network}

We follow the standard VAE procedure to perform approximate inference on the volume transformation parameters. Given the volumetric image $\mathbf{V}\in \mathbb{R}^{3}$, which is a single-channel grayscale volume in a 3D spatial domain, the inference network of our system conjectures corresponding posterior of rotation $\Theta$ (defined in Equation \ref{eqn:rotation}), translation $\Delta \mathbf{t}$ and latent variables $\mathbf{z}$. We represent these approximate posterior as:
\begin{equation}
    \log  q(\Theta, \Delta\mathbf{t}, \mathbf{z}|\mathbf{V}) = \log \mathcal{N}(\Theta, \Delta \mathbf{t}, \mathbf{z}; \mu(\mathbf{V}),\sigma^2(\mathbf{V}) I ) 
    \label{eq1}
\end{equation}
For the choosing of priors on $\mathbf{z}$, we use the usual $\mathcal{N}(0, I)$ prior. As the choosing of appropriate priors on $\Theta$ and $\Delta\mathbf{t}$, we use the Gaussian prior with mean zero, $\mu = 0$. The standard deviation of the Gaussian distribution $\sigma^2$ is set to control the model's tolerance to large volume rotation and translation. The KL divergence penalizes the distance between the chosen distribution priors and the inference network's approximate posterior.

\begin{align}
KL_{\Theta} &=\sum_{\theta \in \Theta} \{- \log{\sigma_\theta} + \log{s_\theta} + \frac{\sigma_\theta ^2 + \mu_\theta^2}{ 2 s_\theta^2} -0.5\}.\\
KL_{ \Delta \mathbf{t}} &=\sum_{t \in  \Delta \mathbf{t}} \{- \log{\sigma_t} + \log{s_t} + \frac{\sigma_t ^2 + \mu_t^2}{ 2 s_t^2} -0.5\}.\\
KL_{\mathbf{z}} &=\sum_{z \in \mathbf{z}} \{ - \log{\sigma_z} +0.5* \sigma_z ^2 + 0.5*\mu_z^2 -0.5\}.
\end{align}

\noindent where $\sigma_\theta, \sigma_t$ and $\sigma_z$ are given by the inference network for volume $\mathbf{V}$. $s_\theta$ and $s_t$ are the standard deviation of the prior on $\Theta$ and $\Delta \mathbf{t}$.

\subsection{3D Spatial Generator Network}
We model the decoder of our system as the function of the 3D spatial coordinate vector, $\mathbf{x}_i \in \mathbf{C}$ , which is the key part of our 3D-SpVAE network. Different from standard VAE, the decoder of our system not only adapt the unstructured latent variables, $\mathbf{z}$, but also the 3D coordinate  $\mathbf{x}_i = (x_{i},y_{i},z_{i})$. We parametrize the decoding as a function of $\mathbf{z}$ and $\mathbf{x}_i$, which infers the probability of observing the voxel intensity, $v_i$, at coordinate, $\mathbf{x}_i$. The distribution, $p_g(v_i|\mathbf{x}_i, \mathbf{z})$ is defined to be Gaussian. The decoder is formulated as a multilayer perceptron (MLP) with parameter $g$.

When given the volume rotation $\Theta$, the translation $\Delta\mathbf{t}$ and the unstructured latent variable $\mathbf{z}$, the spatial generative network infers the probability of observing volume $\mathbf{V}$ as:

\begin{equation}
\log p_g(\mathbf{V}|\Theta,\Delta\mathbf{t},\mathbf{z}) = \sum_{\mathbf{x}_i \in \mathbf{C}} \log p_g(v_i|\mathbf{x}_i R(\Theta) + \Delta\mathbf{t},\mathbf{z}).
\label{eq5}
\end{equation}

Unlike usual image generative models, we learn a function to decode a single voxel probability independently conditioned on the spatial coordinate, rather than reconstruct the entire image as the decoder's output. Our formulation of the volume rotation and translation is differentiable everywhere, enabling the error back-propagation in the end-to-end training of the network.

\subsection{Variation Lower-Bound}
In this section, we consider the variation lower bound of our 3D Spatial-VAE as the original likelihood term can not be calculated directly. The inference network conjectures the approximate posterior of the rotation, $q(\Theta|\mathbf{V}) = \mathcal{N}(\mu_{\Theta}, \sigma_{\Theta}^2(\mathbf{V}))$, the translation, $q(\Delta\mathbf{t}|\mathbf{V}) = \mathcal{N}(\mu_{\Delta\mathbf{t}}, \sigma_{\Delta\mathbf{t}}^2(\mathbf{V}))$ and the unstructured latent variables, $q(\mathbf{z}|\mathbf{V}) = \mathcal{N}(\mu_\mathbf{z}, \sigma_\mathbf{z}^2(\mathbf{V}))$. We denote the variables given by the inference network collectively as $\Omega$. The variational lower-bound of our system balance the terms regarding the volume reconstruction and the distribution approximation.    

\begin{equation}
\label{eq:ELBO}
\mathcal{L} = \mathop{\mathbb{E}}\limits_{\Omega \sim q(\Omega|\mathbf{V})} \log p_g(\mathbf{V}|\Theta,\Delta\mathbf{t},\mathbf{z}) - \text{KL}(q(\Omega|\mathbf{V})||p(\Omega)),
\end{equation}

\begin{equation}  
\begin{aligned}
\text{KL}(q(\Omega|&\mathbf{V})||p(\Omega)) = KL(q(\Theta|\mathbf{V})||p(\Theta)) + \text{KL}(q(\Delta\mathbf{t}|\mathbf{V})||p(\Delta\mathbf{t}))  +  \text{KL}(q(\mathbf{z}|\mathbf{V})||p(\mathbf{z})).
\end{aligned}
\end{equation}

\noindent where $p(\Delta\mathbf{t})= \mathcal{N}(0, s_{\Delta\mathbf{t}}^2)$ and  $p(\mathbf{z})= \mathcal{N}(0, I)$. The first term of $\mathcal{L}$ is to measure the difference of the input volume $\mathbf{V}$ and the reconstructed volume. The second term of $\mathcal{L}$ is to enfore the inferred posteriors to approximate the distribution priors.

\subsection{Volume Transformation Formulation}

We formulate the volume rotation with the Euler angles including \emph{yaw, pitch} and \emph{roll}. The rotation matrix for rotation $\Theta = [\theta_y,\theta_p, \theta_r]$ is as follows:

\begin{equation}
\begin{small}
\begin{array}{lcl}
R(\Theta) =R(\theta_y)R(\theta_p)R(\theta_r) = \\
\left[ 
\begin{array}{ccc}  
\cos(\theta_y) & -\sin(\theta_y) & 0\\ 
\sin(\theta_y) & cos(\theta_y) & 0\\ 
0 & 0 & 1\\ 
\end{array}
\right]
 \left[   
\begin{array}{ccc} 
\cos(\theta_p) & 0 & \sin(\theta_p)\\  
0 & 1 & 0\\  
-\sin(\theta_p) & 0 & \cos(\theta_p)\\  
\end{array}
\right]
 \left[    
\begin{array}{ccc}  
1 & 0 & 0\\  
0 & \cos(\theta_r) & -sin(\theta_r)\\ 
0 &sin(\theta_r) &\cos(\theta_r)\\ 
\end{array}
\right].
\end{array}
\end{small}
\label{eqn:rotation}
\end{equation}

We model the rotations and translations of the volume as the transformation of the volume coordinate space. The 3D spatial coordinate, $\mathbf{C}\in \mathbb{R}^{4}$ is a 4-channel tensor, in which each channel is a volume with the same size as $\mathbf{V}$. Specifically, the translation of the volumetric image in 3D space by $\Delta \mathbf{t}$ corresponds to translate the 3D spatial coordinate $\mathbf{C}$ by  $\Delta \mathbf{t}$. The rotation of the volume by $\Theta$ corresponds to rotate the 3D spatial coordinate $\mathbf{C}$ by $\Theta$.

\section{Experiments}
\subsection{Datasets, Experimental Setting, and Training Algorithm}
To evaluate the performance of our proposed method, we design a series of experiments on simulated datasets with different signal-to-noise (SNR) level and two prevailing real datasets to illustrate whether our method can encode the transformation factor independently from the semantic. e.g., translations and rotations. 

\paragraph*{Simulated Dataset} We use \textit{aitom} platform~\cite{zeng2019aitom} to generate simulated cellular Cryo-ET subtomograms with different SNR levels. To generate raw data with different SNRs, first, we convert protein data bank (PDB) files to density maps. Here we use 1bxn, 1f1b, 1yg6, 2byu, 3gl1, 4d4r, 6t3e, 3hhb as the target object to obtain density maps of every macromolecule and resize to the same size.
Then, we set the simulated box where the single target protein resides with random rotation and location. 
After that, we merge the density map and the box into a hugemap. Finally, we convert hugemap with different SNRs to the subtomogram of the target macromolecule and simulate cryo-ET data. This dataset consists of 8 classes, and each class contains 400 subtomograms with the size of $28^3$ voxels, which have different SNRs. Thus we have 3200 simulate subtomograms in total. We choose SNR 0.01, 0.05, 0.1, respectively, to evaluate the tolerance of our proposed model. We denote this dataset as $\mathcal{S}$.

\paragraph*{Single Particles Real Dataset} We also use the Noble Single Particles Dataset~\cite{noble2018routine} to handle sequential experiments. This dataset consists of 7 classes, and each class contains 400 subtomograms with the size of $28^3$ voxels, which have different scaling factors. Thus we have 2800 real subtomograms in total. These seven structural classes are rabbit muscle aldolase, DNAB  helicase-helicase, hemagglutinin,  T20S  proteasome,  apoferritin,  insulin receptor, and glutamate dehydrogenase.  We denote this dataset as $\mathcal{C}$.
    
\paragraph*{Bacterial RNA Polymerase Real Dataset} This dataset contains 1,566 subtomograms with size $28^3$ extracted from four tomograms of purified bacterial RNA polymerases~\cite{chen2019eliminating}. We denote this dataset as $\mathcal{R}$. Compared with the real dataset $\mathcal{C}$, this dataset $\mathcal{R}$ is a much easier one with fewer classes, more images per class, and lower SNR on average.

\begin{algorithm}
\caption{Training procedure of 3D-SpVAE}
\begin{algorithmic}[1]
\label{algor}
\STATE Input: Given batch size $M$, the batch 3D cryo-ET training dataset $\mathcal{X}_{\text{tr}}=\{V_i\}|_{i=1}^{M}$, epoch number $N$, latent variable dimension $d$, learning rate $\alpha$.
\FOR{epoch = 1, 2, ..., $N$}
\STATE Feed $\mathcal{X}_{\text{tr}}$ to the 3D inference network to get latent variables $\Omega = (\Theta, \Delta t, z)$ by using Eqn.~\ref{eq1}. 
\FOR{$V_i\in \mathcal{X}_{\text{tr}}$}
    \FOR{each 3D spatial coordinate $\boldsymbol{x}_{i}$ in $V_i$}
    \STATE Calculate transformed coordinates $\Tilde{\boldsymbol{x}_{i}}$ using $\Theta$ and $\Delta t$ using transformations in Eqn.~\ref{eq5}.
    \STATE Compute reconstructed volume intensity $\boldsymbol{v}_{i}$ at $\boldsymbol{x}_{i}$ coordinate through 3D spatial generator network using $\Tilde{\boldsymbol{x}_{i}}$ and $z$ mentioned in Eqn.~\ref{eq5}.
    \ENDFOR
\ENDFOR
\STATE Calculate reconstructed loss $\text{ELBO}$ in Eqn.~\ref{eq:ELBO}.
\STATE Calculate gradient $\nabla \mathcal{L}$.
\STATE Update parameters in encoder and decoder $\mathcal{P} \leftarrow \mathcal{P}-\alpha\nabla\mathcal{L}$.
\ENDFOR
\end{algorithmic}   
\end{algorithm}

We train 3D-SpVAE and 3D-SpVAE++ use the ADAM algorithm~\cite{kingma2014adam} with a learning rate of 0.0001 and minibatch size of 128 to train all the models. In testing, we inference on each per 3D cryo-ET data of test set and obtain generated results. Empirically, we set the standard deviation of translation latent variables to 0.14 and the standard deviation on the rotation prior to be $\pi/2$. More details could be found in the algorithm~\ref{algor} and supplementary.

\begin{figure*}[!htbp]
    \centering
    \includegraphics[trim=0 130 0 130, clip, width=1.0\textwidth]{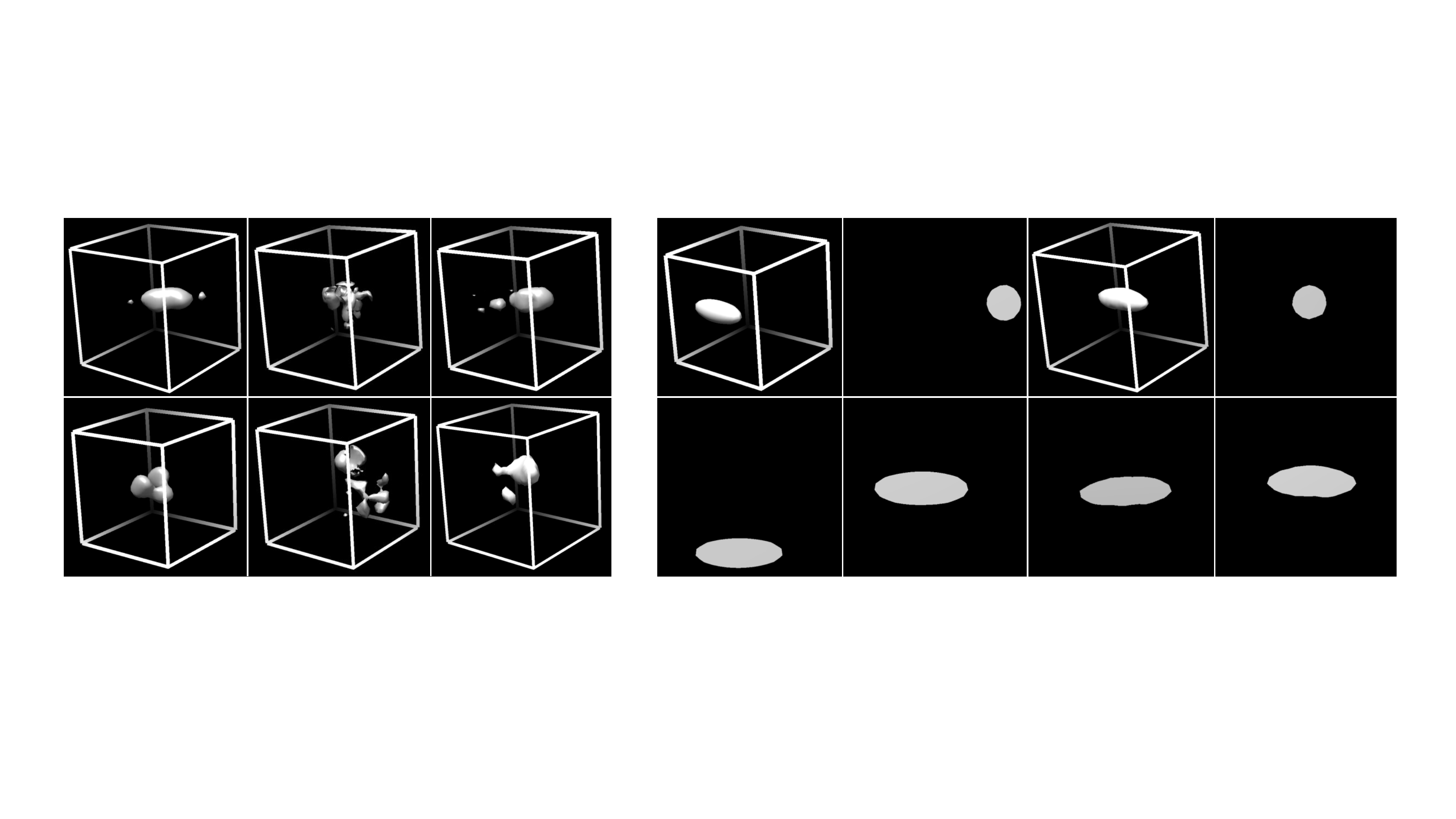}
    \caption{Comparative visualization results of different methods on synthesized and real dataset. \textit{Left}: results on synthesized data. The first column is 3D ground truth, the second are generated results of 3D vanilla VAE and the last are results of 3D-SpVAE. \textit{Right}: results on real dataset. The first two columns are original protein position from left, front, top views by clockwise order and the last two are results of 3D-SpVAE++. From the visualized results, we can observe our proposed 3D-SpVAE and 3D-SpVAE++ can disentangle semantic transformation factors on both synthesized and real cryo-ET data.}\label{fig:syn-vis}
    \label{fig:cryo-et-vis}
\end{figure*}  

\subsection{Evaluations on Synthesized Cryo-ET data}
To evaluate our proposed model's performance, we first experiment on our synthesized cryo-ET data. In real data, the SNRs are always among $[0.01, 0.1]$; thus, we uniformly choose three SNRs and generate synthesized cryo-ET samples. We show our results at Tab.~\ref{tb:syn-results}. We find our proposed 3D-SpVAE and 3D-SpVAE++ models can fit the data distributions better, which leads to significantly lower KL divergence; shown in the third column. Even though we observe the ELBO of our methods are slightly worse than the baseline; shown in the second column, we find the quality of reconstructed subtomograms, which is more important and realistic, are significantly better than the baseline; shown in Fig.~\ref{fig:cryo-et-vis}. This may suggest ELBO may not be a representative matrix to show the disentangle ability, so we will also present the more intuitive latent space traversal experiments to show the disentangled representation power at Sec~\ref{vis_latent_space}. 

\begin{table}[!htbp]
\centering
\resizebox{0.8\textwidth}{!}{
\begin{tabular}{c|ccc|ccc}
\toprule
\multirow{1}{*}{Metric Type}          & \multicolumn{3}{c|}{\multirow{1}{*}{ELBO}} & \multicolumn{3}{c}{\multirow{1}{*}{KL-Divergence}} \\ \midrule
SNR & 0.01 & 0.05 & 0.1 & 0.01 & 0.05 & 0.1\\ \midrule
3D Vanilla VAE & -132.10 & -141.62 & -140.81 & 3.58 & 11.06 & 5.66\\
\textbf{3D-SpVAE} & -263.63 & -260.48 & -184.00 & 0.02 & 0.03 & 0.06\\
\textbf{3D-SpVAE++} & -263.65 & -260.45 & -183.85 & 0.07 & 0.08 & 0.05\\
\bottomrule 
\end{tabular}
}
\caption{ELBO and KL-Divergence on synthesized data with different SNRs.}
\label{tb:syn-results}
\end{table}

\subsection{Evaluation on Cryo-ET Real Datasets}
We also conduct extensive experiments on the above mentioned two real dataset $\mathcal{C}$ and $\mathcal{R}$. Noted that these two datasets have significantly different data distributions. We show our experiments at the second column in Fig.~\ref{tb:DG-EXP}. $\mathcal{C}\to\mathcal{C}$ means we train on real dataset $\mathcal{C}$ and evaluate on $\mathcal{C}$ as well. $\mathcal{R}\to \mathcal{R}$ has the same meaning. If we compare only on the ELBO, we find our methods performs slightly better than the baseline 3D Vanilla VAE.

\begin{table}[!htbp]
\centering
\resizebox{1.0\textwidth}{!}{%
\begin{tabular}{c|c|cc|cc|cc}
\toprule
Metric & $\mathcal{S}\to\mathcal{S}$ & $\mathcal{R}\to\mathcal{R}$ & $\mathcal{C}\to\mathcal{C}$ & $\mathcal{S}\to \mathcal{R}$ & $\mathcal{S}\to \mathcal{C}$ & $\mathcal{C}\to \mathcal{R}$ & $\mathcal{R}\to \mathcal{C}$\\ \midrule
3D Vanilla VAE & -140.81 & -43.95 & -726.44 & -139.95 &  -3316.24 & -2626.83 & -3288.74\\
\textbf{3D-SpVAE} & -184.00 & -41.90 & -712.70 & \textbf{-41.87} & -3300.88 & \textbf{-2581.18} & \textbf{-3273.15}\\
\textbf{3D-SpVAE++} & -183.85 & -39.45 & -690.74 & \textbf{-42.75} &\textbf{-3270.70}  & -2610.25 & \textbf{-3273.17}\\
\bottomrule
\end{tabular}%
}
\caption{ELBO performance. \textit{Left}: ELBO on synthetic and real cryo-ET datasets. \textit{Right}: cross-domain evaluations.}
\label{tb:DG-EXP}
\end{table} 

\subsection{Cross-Domain Evaluation}
We consider domain generalization (DG), where the training data and testing data are from significantly different domains. As previously described, we've done a series of experiments on synthesized cryo-ET datasets and two real datasets with significantly different distributions. To show whether our proposed methods can be generalized to disentangle transformation factors with data from a different domain, we focus on investigating the following things in this section:

\begin{itemize}
    \item \textit{Q1: Is synthesized data well enough to represent essential features of real data?} We first train several models on synthesized data $\mathcal{S}$ and then transfer the learned model to solve the explicitly disentangled representation problem on real cryo-ET datasets.
    \item \textit{Q2: Is our model stable enough to transfer knowledge among real cryo-ET datasets?} According to the general domain generalization setting, we push our model to learn from source ${\mathcal{C}|\mathcal{R}}$ and transfer the knowledge to another target domain ${\mathcal{R}|\mathcal{C}}$.
\end{itemize}

We first investigate on \textit{Q1} by transfering the learned model on synthetic dataset $\mathcal{S}$ to different real datasets, \textit{i.e.,} $\mathcal{R}$ and $\mathcal{C}$, the results are shown in the fourth column module in Tab.~\ref{tb:DG-EXP}. If we compare the results of $\mathcal{S}\to \mathcal{S}$ and $\mathcal{S}\to \mathcal{R}$, we find our method can be transferred well to real data distribution while being trained on the synthetic dataset only. Taking the 3D-SpVAE++ as an example, we achieved $-42.75$ ELBO on $\mathcal{S}\to \mathcal{R}$, while the 3D Vanilla VAE can only achieve $-139.95$, which is significantly lower than ours. Besides, we're surprised to find the transfer learning ability is significant as we achieved better ELBO on real dataset with the model trained on the synthesized dataset (\textit{i.e.,} -42.75 vs. -183.85). We also conduct the cross-domain evaluations with the much difficult $\mathcal{C}$ real dataset. We found even though our methods show better ELBO than the baseline, all methods failed to generalized well to this hard dataset. We assume this is because there is an extreme data distribution gap between $\mathcal{S}$ and $\mathcal{C}$.
   
We also conduct experiments to answer \textit{Q2} about model transferability among real datasets.  We show the comparative results at the last column module in Tab.~\ref{tb:DG-EXP}. We also show better results than the baseline model, which means our model works better transferring knowledge among real cryo-ET datasets.

\subsection{Visualizations of Latent Space}\label{vis_latent_space}
Inspired by the analysis of latent space in $\beta-\text{VAE}$, we do latent traversal for six structured latent variables of translation and rotation. Three of them describe the rotation and the other three describe the translation in three dimensions. Given a specific 3D protein structure, we first obtain the initial latent representation through the inference network, then traverse values of a single latent variable within a proper range for rotation (in [$-2$, $2]$) or translation (in [$-0.75$, $0.75$]) while fixing the remains, and plot the reconstruction results through the spatial generator network.

\begin{figure}[!htbp]
    \centering
    \includegraphics[width=1.0\textwidth, trim={80 160 80 160}, clip]{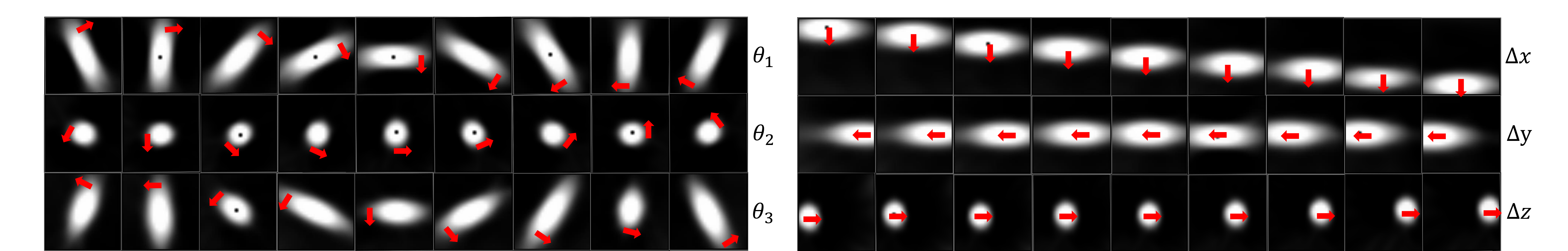}
    \caption{Semantic transformation factor disentangling and latent traversal visualization. Left for rotation, right for translation and red arrows indicate the direction of rotation or translation of the same point on the 2D projection plane.}
    \label{fig:traversal}
\end{figure}

The process we observe and show the latent traversal visualization is shown schematically in the supplementary. Latent traversal visualization results can be seen in Fig.~\ref{fig:traversal}, each row contains 9 2D projection planes corresponding to the traversal of a single latent variable. The difference value is 0.5 of rotation and 0.1875 of translation between two snaps.

The disentangling and traversal are evident: different values of the latent variable independently code for different rotation angles and translation distances. The left 3 lines represent rotation angle changes. E.g. the first row shows a trend of uniform clockwise rotation as $\theta_1$ gradually increases, and the protein does a half cycle rotation from the second snap to the eighth snap so that we could find an rotation latent value changes by 0.5 corresponding to 30 degrees approximately in our model. The right 3 lines represent translation distance changes. For example, the first row shows a trend of top-down translation as $\Delta x$ gradually increases. The protein moves almost across the whole snap during all snaps, thus translation latent value changes by 0.1875 corresponding to 3.5 pixels approximately here.

\section{Discussion}  
In this paper, we proposed a novel 3D-SpVAE to explicitly disentangle transformation semantic factors including translations and rotations. We show strong qualitative and quantitative results on both synthesized datasets with different SNRs and real datasets. Cross-domain evaluations show our method is promising in transfer learning among datasets with different distributions. Our traversal experiments prove 3D-SpVAE can disentangle semantic transformations from unstructured factors. We believe the idea in this paper is promising and can be potentially very useful to understanding macromolecular structures captured by cryo-ET.

\bibliographystyle{unsrtnat}
\bibliography{references}  

\section*{Supplementary}
\subsection*{Implementation Details and Network Architecture}
We set the dimension of unstructured latent variable to be 50. As we work on 3D cryo-ET data, we choose the dimension of the structured semantic latent variables to be 3 for rotations and translations, respectively.
We fairly train all models with 400 epochs, and we save interval checkpoint after every 10 epochs for evaluation. Models are all implemented by PyTorch and trained on 1 RTX 2080Ti.

We show our network architecture of the baseline method and several variance of our proposed 3D-SpVAE in this section. More details could refer to Tab.~\ref{tb:3D-SpVAE}, \ref{tb:3D-Vanilla-VAE} and \ref{tb:3D-SpVAE++}. Noticed we use the same encoder architecture of different methods.

\begin{table}[!htbp]
\centering
\begin{tabular}{l|l}
\toprule
\multicolumn{1}{c|}{Encoder} & \multicolumn{1}{c}{Decoder} \\ \midrule
Linear(21952,1000), Tanh     & coord\_linear(3,1250)       \\
Linear(1000,1000), Tanh      & latent\_linear(50,1250)     \\
Linear(1000,112)             & Add(1250)                   \\
                             & Reshape(21952,1250), Tanh   \\
                             & Linear(1250,1250), Tanh     \\
                             & Linear(1250,1)              \\
                             & Reshape(1,21952,1)  \\
\bottomrule 
\end{tabular}
\caption{Encoder and Decoder architectures for 3D-SpVAE.}
\label{tb:3D-SpVAE}
\end{table}

\begin{table}[H]
\centering
\begin{tabular}{l}
\toprule
\multicolumn{1}{c}{Decoder} \\ \midrule
Linear(50,1250), Tanh       \\
Linear(1250,1250), Tanh     \\
Linear(1250,21952)          \\
Reshape(1,21952,1)         \\
\bottomrule 
\end{tabular}
\caption{Decoder architectures for 3D Vanilla VAE.}
\label{tb:3D-Vanilla-VAE}
\end{table}

\begin{table}[H]
\centering
\begin{tabular}{l}
\toprule
\multicolumn{1}{c}{Decoder} \\ \midrule
coord\_linear(3,1250)       \\
latent\_linear(50,1250)     \\
Add(1250)                   \\
Reshape(1,1250,28,28,28)         \\
TCONV(1250,128), Relu, BN   \\
TCONV(128,64), Relu, BN     \\
TCONV(64,32), Relu, BN      \\
TCONV(32,1), Relu, BN       \\
Reshape(1,21952,1)      \\
\bottomrule 
\end{tabular}
\caption{Decoder architectures for 3D-SpVAE++.}
\label{tb:3D-SpVAE++}
\end{table} 

\subsection*{Illustration of 3D Projections}
\begin{figure}[!htbp]
    \centering
    \includegraphics[width=0.45\textwidth]{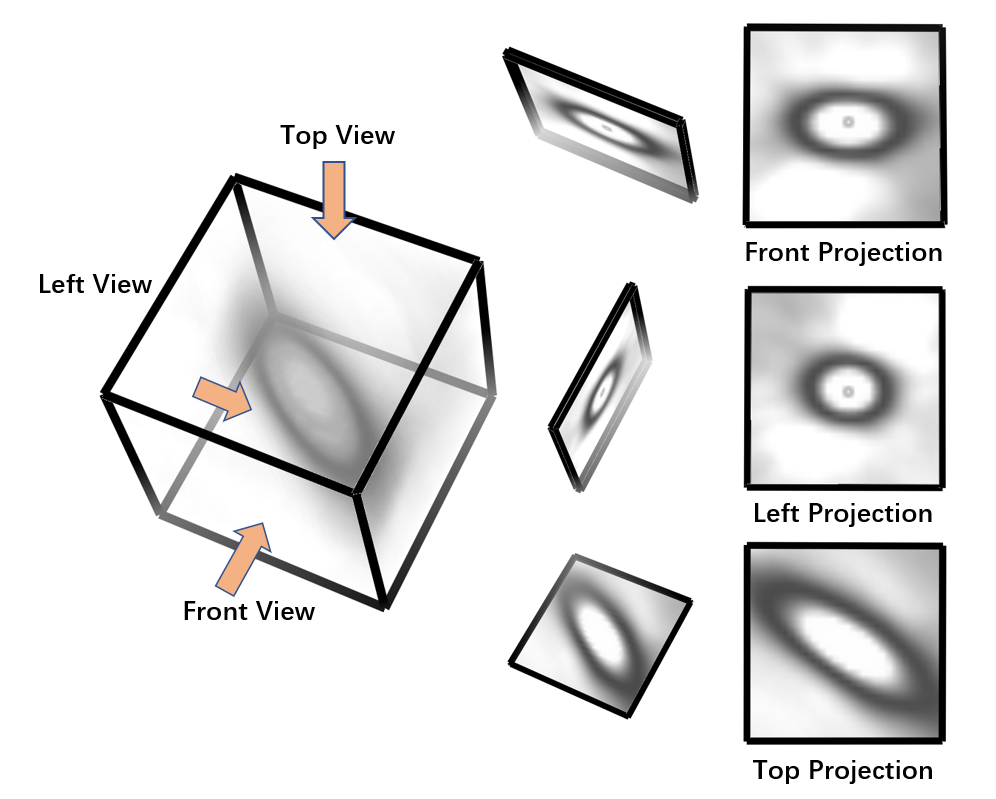}
    \caption{Diagram of 3D projections.}
    \label{fig:diagram_pro}
\end{figure}
Here we illustrate how we project the 3D generated samples to different 2D views (corresponding to the right part of Fig. 2 and Fig. 3 in the main text) in Fig.~\ref{fig:diagram_pro}. For a 3D box containing a reconstructed cryo-ET structure, a 2D projection plane is observed from the top view, the left view, and the front view.
\end{document}